\documentclass[aip,jcp,10pt,superscriptaddress,twocolumn,showpacs]{revtex4-1}

\usepackage{graphicx}
\usepackage{dcolumn}
\usepackage{bm}
\usepackage{amssymb}
\usepackage{url}
\usepackage{natbib}
\usepackage{color}
\usepackage[T1]{fontenc}
\usepackage[utf8]{inputenc}
\usepackage[normalem]{ulem}
\usepackage{subfigure}

\usepackage{graphicx}
\usepackage{amsmath}

\usepackage[colorlinks=true, linkcolor=blue, citecolor=blue,
filecolor=blue, urlcolor=blue, bookmarks=true, bookmarksopen=true,
bookmarksopenlevel=3, plainpages=false, pdfpagelabels=true]{hyperref}

\begin{document}

\title{Reaction energetics of Hydrogen on Si(100) surface: A periodic many-electron theory study}

\author{Theodoros~Tsatsoulis}
\affiliation{Max Planck Institute for Solid State Research, Heisenbergstrasse 1, 70569 Stuttgart, Germany}
\affiliation{Institute for Theoretical Physics, Vienna University of Technology, Wiedner Hauptstrasse 8-10, 1040, Vienna, Austria}
\author{Sung~Sakong}
\affiliation{Institute of Theoretical Chemistry, Ulm University, Albert-Einstein-Allee 11, 89081 Ulm, Germany}
\author{Axel~Gro{\ss}}
\affiliation{Institute of Theoretical Chemistry, Ulm University, Albert-Einstein-Allee 11, 89081 Ulm, Germany}
\author{Andreas~Gr{\"u}neis}
\email{andreas.grueneis@tuwien.ac.at}
\affiliation{Max Planck Institute for Solid State Research, Heisenbergstrasse 1, 70569 Stuttgart, Germany}
\affiliation{Institute for Theoretical Physics, Vienna University of Technology, Wiedner Hauptstrasse 8-10, 1040, Vienna, Austria}

\date{\today}

\begin{abstract}
We report on a many-electron wavefunction theory study for the
reaction energetics of hydrogen dissociation on the Si(100)
surface. We demonstrate that quantum chemical wavefunction based
methods using periodic boundary conditions can predict chemically
accurate results for the activation barrier and the chemisorption
energy in agreement with experimental findings. These highly accurate
results for the reaction energetics enable a deeper understanding of
the underlying physical mechanism and make it possible to benchmark
widely used density functional theory methods.
\end{abstract}

\maketitle

\section{Introduction}

Reactions of gas-phase molecules on surfaces
play an important role in many physical and chemical processes.  Along
a surface reaction path, the energetics of the combined
molecule--surface system varies significantly according to the
rearranged chemical bonds featuring charge transfer, covalent bonding,
and weak van der Waals interactions. Arrhenius relation indicates a
small error in the activation energy can cause a large change of the
reaction rate. Thus a many-electron theory able to describe a wide
range of exchange and correlation effects in molecules, solids, and
molecule--surface systems simultaneously is required to estimate a
reaction scheme with chemical accuracy~(1~kcal/mol).  The
computational method of choice for such problems is density functional
theory (DFT) due to its good trade-off between accuracy and
computational cost~\cite{kohn1965,kroes1999,gross2014}. However,
several shortcomings exist in the most widely used approximate
exchange-correlation functionals~\cite{cohen2012}. Density functionals
based on parametrizations that achieve accurate results only for
either solids or gas-phase molecules, introduce systematic errors in
combined molecule--surface systems~\cite{schimka2010}.  Many local and
semi-local density functional approximations underestimate reaction
barriers, often due to the self-interaction
error~\cite{perdew1981,cohen2012}. The lack of chemically accurate
benchmark results for molecule--surface systems limits our
understanding of the various origins of error in currently available
density functionals.  High level \emph{ab-initio} wavefunction
theories, such as the coupled cluster (CC) method, predict molecular
reactions as well as properties of solids with chemical
accuracy~\cite{Helgie,zheng2009,booth2013,yang2014}.  However, owing
to their large computational cost these methods have so far only been
applied to small cluster models or used in embedding techniques when
applied to
surfaces~\cite{voloshina2011,libisch2012,kubas2016,boese2016}.  A
careful validation against periodic high level wavefunction methods is
still missing to date.

In this work we consider a prototypical molecule--surface reaction:
the dissociative adsorption of molecular hydrogen on the Si(100)
surface~\cite{hoefer1992,pehlke1995,radeke1996,gross1997,biedermann1999,penev1999,duerr2001,raschke2001,steckel2001,filippi2002,duerr2006,brenig2008}.
Previous studies identify two reaction paths of dissociative H$_2$
adsorption, termed the intra- (H$2^{\ast}$) and inter-dimer (H2)
pathways, as shown in Fig.~\ref{fig:poscarTransition}.  Along both
reaction paths, the stretch of H--H bond is accompanied by a
significant modification of the characteristic buckled Si-dimer
configuration in the vicinity of the
molecule~\cite{wolkow1992,ramstad1995,healy2001,brenig2008}.  These
structural modifications induce delicate changes to the electronic
structure.  DFT methods based on generalized gradient approximation
(GGA) capture the changes in the electronic exchange and correlation
effects poorly along the reaction paths and result in too small
adsorption barriers and reaction energies compared to
experiments~\cite{pehlke1995,penev1999,duerr2001}.  Quantum Monte
Carlo (QMC) and quantum chemistry methods using finite clusters
predict adequate adsorption barriers for both pathways, however,
reaction energies are
overestimated~\cite{filippi2002,radeke1996,steckel2001}.  In previous
DFT and high level correlated calculations, H$_2$ adsorption is
hindered by the smallest barrier through the H2 pathway, and the
reaction occurs via a pairing mechanism~\cite{duerr2006,brenig2008}.

Here, we present a periodic quantum chemical description of the
reaction using a recently implemented periodic CC theory, applicable
to molecule--surface
systems~\cite{hummel2017,tsatsoulis2017,gruber2018}.  We show that
activation and reaction energies are calculated to within chemical
accuracy compared to experimental values. Most interestingly, it
becomes clear that the adsorption barriers for the H$2^{\ast}$ and H2
pathways are very similar, in contrast to previous findings.  We will
demonstrate that the main source of error of DFT-GGA is the self
interaction error leading to an incorrect ground state density for the
H2 path.

\section{Computational details}

\begin{figure}
  \includegraphics[width=7cm]{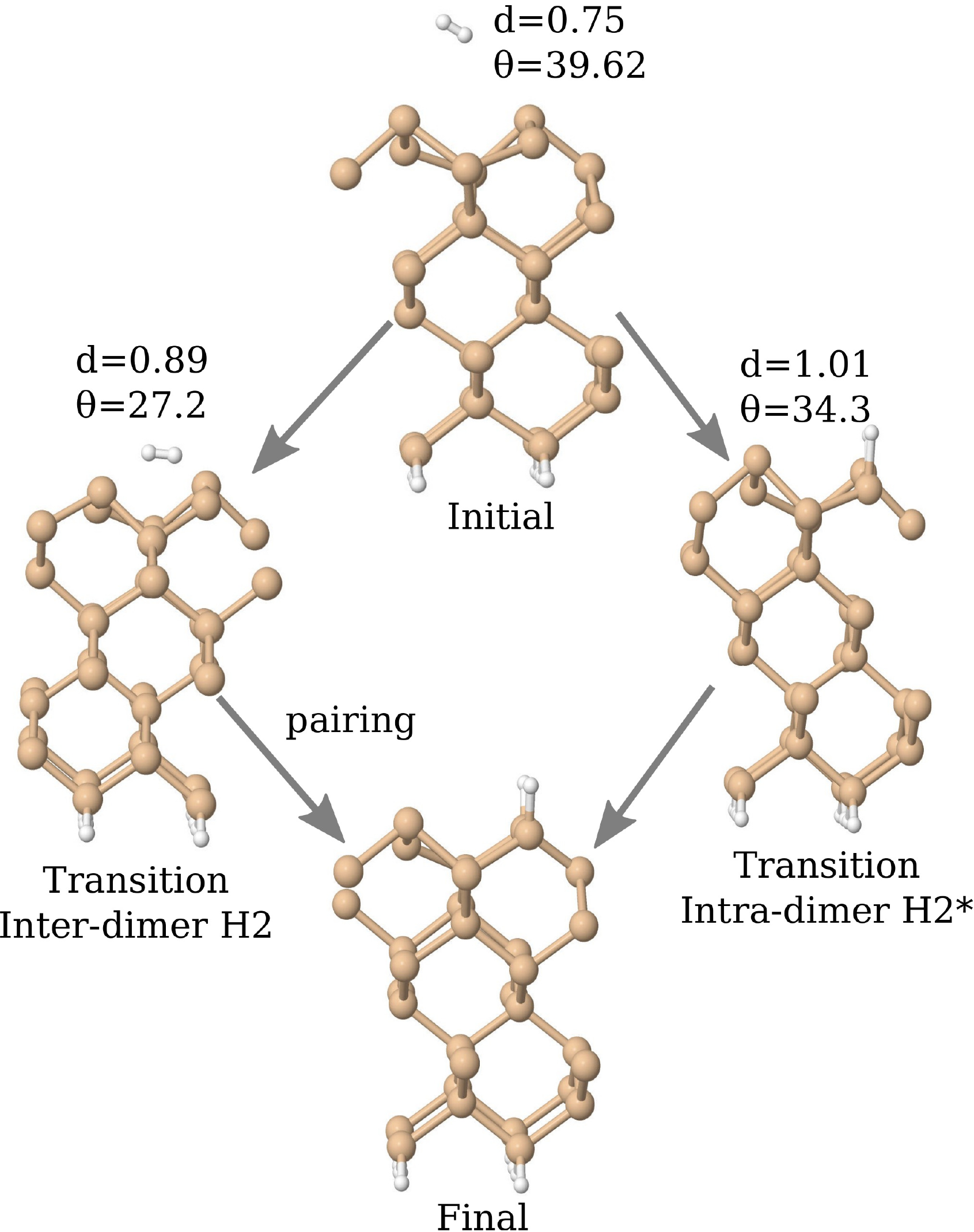}
  \caption{Intra- (H$2^{\ast}$) and inter-dimer (H2) reaction pathways at low
    coverage (one H$_2$ molecule per two Si dimers). $d$ denotes the
    bond length of the H$_2$ molecule in {\AA} whereas $\theta$ the
    buckling angle of the Si dimers in degree.}
  \label{fig:poscarTransition}
\end{figure}

We employ periodic slabs for all density-functional and wavefunction
based calculations. A Si(100)-$2\times2$ surface with 8-layers is
used, terminated with hydrogen atoms to passivate dangling bonds at
the bottom layer.  All calculations involve a plane-wave basis within
the full potential projector-augmented-wave method (PAW) as
implemented in the {\sc VASP}
code~\cite{kresse1994,kresse19962,kresse1996,bloechl1994,kresse1999}.  In all
calculations the $1s$ electronic states of the H atoms and the $3s$
and $3p$ states of the Si atoms were treated as valence states.  The
minimum energy paths of the H$2^{\ast}$ and H2 reactions are
determined using the nudged elastic band (NEB) method within
variational transition state theory~\cite{henkelman9978}.  We used 8
images for the calculations. For the electronic structure calculations
we used the Perdew-Burke-Ernzerhorf (PBE) exchange correlation
functional~\cite{perdew1996}. One-electron states were expanded using
a plane-wave basis with a cutoff energy of 250~eV, alongside an
$8\times8\times1$ $k$-mesh to sample the first Brillouin zone.  The
exact energies of the transition states are determined by an
interpolation or by using the the climbing NEB
method~\cite{henkelman9901}.  The corresponding structures are shown
in Fig.~\ref{fig:poscarTransition}.  We use the same geometries for
the initial, transition, and final states to determine the reaction
energy and the adsorption and desorption barriers for all
methods. This allows for a direct comparison of the different levels
of theory. Contributions of vibrational zero-point energies (ZPE) are
included in all calculations and were taken from
Ref.~\onlinecite{steckel2001}.  A $4\times4\times1$ $k$-mesh was
employed for all DFT calculations, while the plane wave energy cutoff
was set to 500~eV. We have explored the accuracy of several
density-functional approximations covering all five rungs of the
Jacob's ladder of DFT proposed by Perdew and
Schmidt~\cite{perdew2001}.  Periodic wavefunction based calculations
were also performed using a plane-wave basis within the PAW framework
using the {\sc VASP} code and an interface to the coupled cluster code
{\sc CC4S} employing the Cyclops Tensor Framework ({\sc
CTF})~\cite{solomonik2014}.  Hartree--Fock (HF) calculations were
converged within the plane-wave basis.  For second order perturbation
theory (MP2), CC singles and doubles (CCSD) and perturbative triples
(CCSD(T)), as well as the random phase approximation (RPA), we employ
a set of atom-centered Gaussian-type functions based on Dunning's
correlation consistent polarized Valence Quadruple Zeta basis set
augmented with diffuse functions
(aVQZ)~\cite{dunning1989,feller1996,schuchardt2007}, mapped onto a
plane-wave representation~\cite{booth2016}, to construct the
unoccupied one-electron states.  We always rediagonalize the Fock
matrix in order to perform a canonical correlated calculation.  For
the CCSD and (T) calculations we further reduce the number of
unoccupied states using MP2 natural orbitals (NOs), obtained from the
virtual--virtual orbital block diagonalization of the one-electron
reduced density matrix at the level of MP2~\cite{grueneis2011}. We
always correct for the remaining basis set incompleteness error using
an estimation based on a $\Gamma$-point full plane-wave direct MP2
calculation. A $4\times4\times1$ $k$-mesh was employed for the
twist-averaging procedure~\cite{gruber2018} used in the CCSD
calculations, whereas the remaining finite size error of the
correlation energy was corrected for using an interpolation technique
of the structure factor on a plane-wave grid~\cite{gruber2018}.
CCSD(T) results were obtained as correction to CCSD using the
$\Gamma$-point approximation.

\section{Results and discussion}

\begin{figure}
  \includegraphics[width=8.5cm]{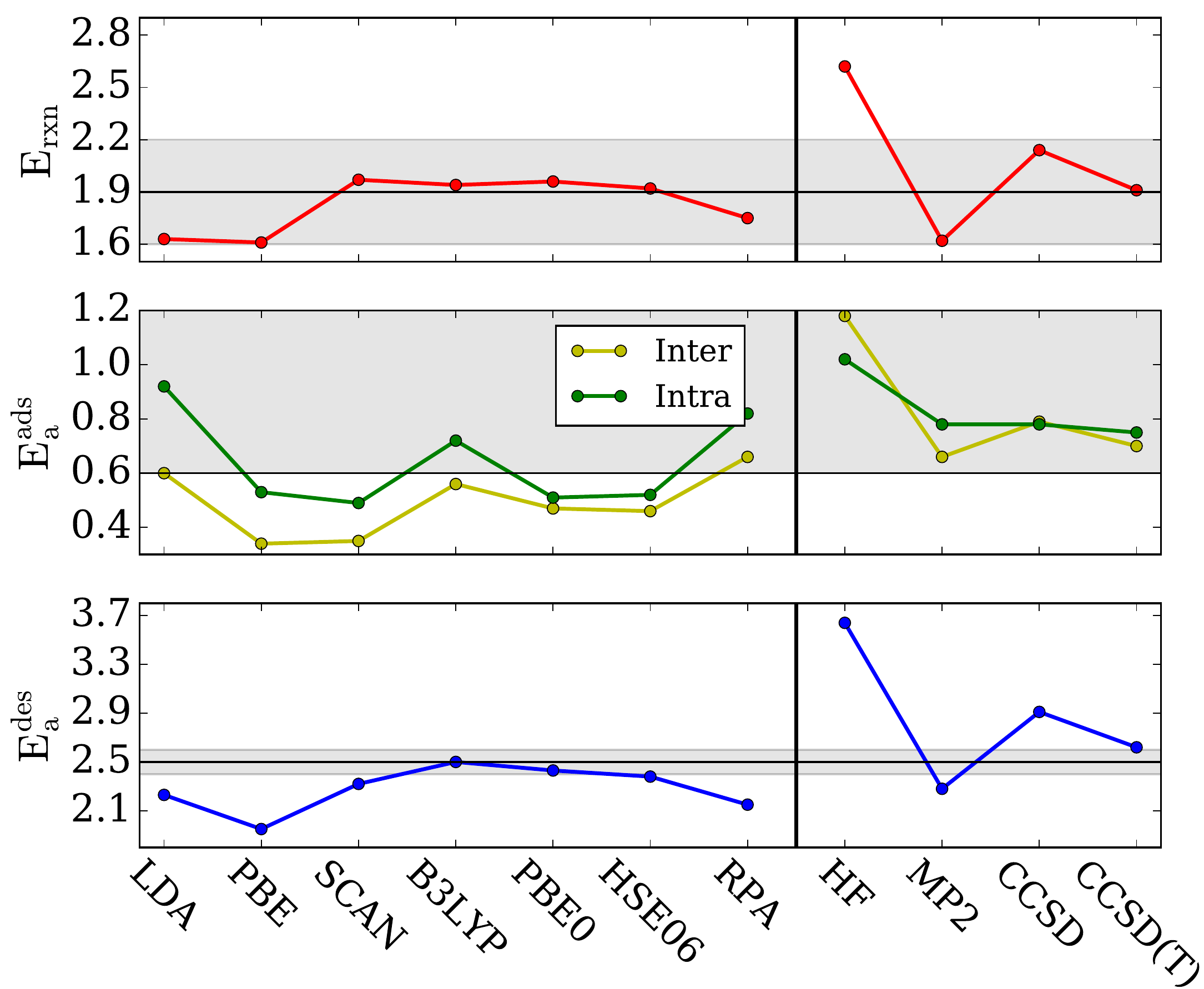}
  \caption{Reaction energetics for H$_2$ dissociation on the Si(001)
    surface calculated at different levels of theory. The calculated
    reaction energies ($\mathrm{E}_{\mathrm{rxn}}$) adsorption
    ($\mathrm{E}^{\mathrm{ads}}_{\mathrm{a}}$) and desorption
    ($\mathrm{E}^{\mathrm{des}}_{\mathrm{a}}$) barriers for H$_2$
    dissociative adsorption on Si(100) surface. The black lines
    represent the experimental estimate while the shaded region the
    error. All energies given in eV.}
  \label{fig:energy}
\end{figure}

Figure~\ref{fig:energy} shows the calculated reaction
energetics for H$_2$ on the (100) surface of Si at different levels of
theory, together with the experimental estimates and the corresponding
error depicted by the shaded area. The results are also summarized in
Table.~\ref{tbl:transition}. We first consider the reaction energy
$(\mathrm{E}_{\mathrm{rxn}})$ shown in the top panel and we will turn
to the discussion of barrier heights later. The reaction energy is
defined as $\mathrm{E}_{\mathrm{rxn}} = \mathrm{E}_{\mathrm{initial}}
- \mathrm{E}_{\mathrm{final}}$, where the corresponding final and
initial structures are shown in Fig.~\ref{fig:poscarTransition}.  The
experimental estimate for the reaction energy is
1.9$\pm$0.3~eV~\cite{raschke2001}.  The local-density approximation
(LDA) constitutes the lowest rung of Perdew's Jacob's ladder of DFT
methods.  Reaction energies computed in the LDA are approximately
0.3~eV too small compared to the experimental estimate.  The PBE
functional is one of the most extensively used GGA functionals which
represent the second rung of the Jacob's ladder and noticeably
underestimates the reaction energy without any improvement compared to
LDA. The meta-GGA functionals lie on the third rung and utilize the
kinetic energy together with the electron density and its
gradient. The strongly constrained and appropriately normed
(SCAN)~\cite{sun2015,sun2016} meta-GGA density functional
significantly improves the reaction energy ($\mathrm{E}^{\rm
  SCAN}_{\textrm{rxn}}=$1.97~eV), demonstrating its ability to
describe diversely bonded molecules and materials such as the H--Si
system accurately.  Hybrid GGAs provide an improved description of
covalent, hydrogen and ionic bonding by mixing non-local exact
exchange with GGA exchange. B3LYP~\cite{becke1993},
PBE0~\cite{adamo1999}, and HSE06~\cite{krukau2006} yield a reaction
energy with a similar accuracy as SCAN.  The good agreement between
the hybrids and the SCAN functional confirms that meta-GGAs can yield
reaction energies at the same level of accuracy yet at lower
computational cost.  As a method of the fifth rung of Perdew's Jacob's
ladder we examine the RPA. The RPA correlation energy is fully
non-local and seamlessly includes electronic screening as well as
long-range dispersion interactions~\cite{harl2010,ren2012}. The
chemisorption energy in the RPA is, however, underestimated compared
to hybrid-GGA and meta-GGA functionals, in agreement with a well known
underestimation of binding energies~\cite{furche2001}.  Overall we
find that the predicted DFT results for the reaction energy are
improving as one moves from lower to higher rungs with the exception
of the RPA.  However, we attribute the underestimated RPA reaction
energy to the neglect of post-RPA corrrections and a lack of
self-consistency.

We now switch from DFT to the wavefunction based hierarchy for
treating electronic correlation. HF theory, approximating the
many-electron wavefunction by a single Slater determinant,
overestimates the reaction energy by as much as 0.7~eV compared to
experiment.  In passing we note that this is in contrast to
atomization energies of molecules and cohesive energies of solids,
which are usually underestimated by
HF~\cite{helgaker2004,grueneis2010}.  Adding correlation effects at
the level of MP2 theory over-corrects HF and yields a reaction energy
of 1.62~eV, almost 0.3~eV smaller than the experimental estimate. We
assign this overcorrection of MP2 to the small band gap of the Si
surface. The more sophisticated CCSD theory overestimates the
experimental reaction energy by 0.25~eV. Adding the perturbative
triples correction (T) to CCSD yields a reaction energy that is very
close to hybrid DFT results and the experimental estimate.  This
demonstrates the ability of the wavefunction based hierarchy to yield
systematically improvable and chemically accurate chemisorption
energies for molecules on periodic surfaces. However, we note that at
lower levels of theory, the DFT based methods exhibit a significantly
better trade-off between accuracy and computational cost.

Having confirmed experimental measurements for the chemisorption
energy using accurate electronic structure theories, we now seek to
discuss the activation barrier height for the dissociation, which is
defined by $\mathrm{E}^{\mathrm{ads}}_{\textrm{a}} =
\mathrm{E}_{\textrm{transition}} - \mathrm{E}_{\textrm{initial}}$.
Establishing accurate estimates of barrier heights is more difficult
compared to reaction energies for theory as well as experiment.
Transition states often exhibit strong electronic correlation effects
that can only be treated accurately using higher levels of theory.
Furthermore experimental measurements of adsorption barriers are
usually lower bounds and do not allow to determine directly whether
the reaction proceeds via the H2 or H$2^{\ast}$ mechanism.  Adsorption
barriers for both pathways are depicted in the middle panel of
Fig.~\ref{fig:energy}, alongside the experimental lower bound of
0.6~eV~\cite{duerr2001}. LDA yields a barrier of 0.6~eV and 0.92~eV
for the H2 and H$2^{\ast}$ pathway, respectively. The difference in
the barriers between the two pathways is considerable. Noteworthy, LDA
does not vastly underestimate the activation energies, but yields
rather adequetly high barriers for both mechanisms. PBE underestimates
the reaction barriers and yields in agreement with LDA a larger
barrier for the H$2^{\ast}$ pathway. We observe the same trend for the
SCAN functional. Although the description of the reaction energy is
much improved, SCAN fails to ameliorate the errors in the reaction
barriers predicted by GGA, yielding a too low barrier for the H2 path
and the same difference between the barriers of the two
pathways. Interestingly, LDA predicts much larger activation energies
than GGA, challenging the trend LDA$<$GGA observed for adsorption
barriers of molecular reactions~\cite{porezag1995}. Similar results,
however, have been reported for activation energies of gas-phase
reactions~\cite{mahler2017}. Hybrid functionals mix exact exchange
with commonly-used density functionals and partly cancel the spurious
self-interaction error. In the case of H$_2$ on Si(100) hybrid
functionals do improve the description of the reaction barrier.  B3LYP
yields a barrier height of 0.56~eV for the H2 path, whereas PBE0 and
HSE06 yield barriers of 0.47~eV and 0.46~eV respectively. For the
H$2^{\ast}$ mechanism B3LYP predicts a barrier 0.16~eV higher than the
H2 one, whereas PBE0 and HSE06 yield barriers only 0.04 and 0.06~eV
higher than the H2 pathway, respectively. The RPA yields significantly
higher barriers compared to PBE when combined with exact exchange
computed with PBE orbitals. However, we note that the H2 path is still
favored by RPA with a barrier of 0.66~eV compared to 0.82~eV of the
H$2^{\ast}$ path.  The results discussed above illustrate convincingly
a lack of systematic improvability in the obtained estimates of
barrier heights as one moves from lower to higher levels of
approximate DFT based methods. Furthermore, errors in activation
energies may vary sigificantly with the employed density
functional. Therefore reliable predictions for the barrier height and
the relative stability of the considered transition states are not
possible. Nevertheless, the considered system provides a realistic and
insightful scenario to further develop and improve upon the
computationally efficient DFT based methods.

We now turn to the discussion of wavefunction based \emph{ab-initio}
calculations for the barrier. HF theory yields barriers larger than
1~eV for both pathways. In contrast to DFT based findings, HF favors
the H$2^{\ast}$ path over the H2 one by 0.16~eV. In order to better
understand the difference between HF and DFT, we consider the
different paths as a competition between stretching the H$_2$ molecule
and flattening the Si dimers of the surface. In
Fig.~\ref{fig:poscarTransition} the hydrogen bond in the H2 transition
state is 0.89~{\AA}, compared to 1.01~{\AA} of the H$2^{\ast}$ one,
whereas the dimers buckling angle is 27.2$^{\circ}$ for the former and
34.3$^{\circ}$ for the latter transition state. In order to see why HF
favors the H$2^{\ast}$ path we need to consider the energy cost
between the buckled and symmetric configurations of the Si dimer
reconstruction. This energy difference per dimer is 250--260~meV for
LDA and GGA, in contrast to 544~meV for HF. We identify the energy
penalty for flattening the dimers as the main difference between HF
and DFT methods, owing to a large extent on the metallic nature of the
symmetric Si dimer configuration. MP2 theory reverses the preference
of the two pathways. Furthermore, the MP2 adsorption barrier for the
H2 transition state is 0.66~eV while for the H$2^{\ast}$ is
0.78~eV. Due to the smaller band gap of the H2 transition state MP2
overcorrects HF, hence it favors the lower band gap H2 reaction
pathway. CCSD theory yields barriers for the two reaction mechanisms
that are practically degenerate. Specifically, the H2 transition state
barrier is 0.79~eV and the H$2^{\ast}$ 0.78~eV. In agreement with
experiment the inclusion of perturbative triples, CCSD(T), yields
activation barriers of 0.70~eV for the H2 transition and 0.75~eV for
the H$2^{\ast}$ one, retaining the picture of two approximately
degenerate barriers of CCSD to within chemical accuracy.

The picture emerging from the results discussed above is qualitatively
different within the methods we examine. The two barriers are
approximately degenerate using the more sophisticated CCSD and CCSD(T)
theories, in contrast to LDA, PBE and SCAN functionals, where the H2
path is favored. Hybrid functionals remedy partly the self-interaction
error and thus yield barriers that differ less than the GGA and LDA
ones. An exception is B3LYP, where although the barriers are higher in
energy, the H2 path is favored by 0.16~eV. The reason is that part of
the exchange-correlation functional is based on a mixture of LDA
and GGA rather than solely on GGA as in PBE0 and HSE06. Thus B3LYP
contains part of the LDA errors and deficiencies, hence the higher
barriers and the larger difference between the two pathways. Barriers
for the two mechanisms based on the RPA also differ significantly. We
associate the discrepancy between the CCSD(T) barriers and the RPA
ones with the use of PBE orbitals for the RPA calculations as opposed
to the HF ones for CCSD(T). It is likely that a sizeable fraction of
the error exists already in the original DFT functional, leading to an
overestimation of the H2 barrier.  In order to get more insight into
the disagreement of CC methods and DFT based methods, we
performed non-self-consistent calculations for the activation barriers
of the two mechanisms at the level of DFT-PBE using HF orbitals. The
results are shown in Table.~\ref{tbl:transition}. We observe that when
HF orbitals are employed for DFT-PBE calculations the H2 barrier is
appreciably higher, whereas the H$2^{\ast}$ one remains almost the
same. The difference between the two barriers is 0.04~eV in close
agreement with the accurate CCSD(T) and hybrid DFT results. This is
partly due to the cancellation of the density driven one-electron
self-interaction error~\cite{kim2013}, and similar results have been
obtained for simple molecular reaction barriers and adsorption
energies~\cite{janesko2008,patra2018}.

\begin{table}
  \caption{Adsorption barriers for the two pathways, alongside
    desorption and reaction energies. Desorption energies correspond
    to the energetically lowest path, whereas reaction energies to the
    intra-dimer (H$2^{\ast}$) geometry, since it is energetically the lowest
    configuration. ZPE corrections assumed for all
    calculations(identical for both pathways). All energies are
    reported in eV.}
  \label{tbl:transition}
  \begin{ruledtabular}
  \begin{tabular}{lrrrr}
    & \multicolumn{1}{r}{$\mathrm{E}^{\mathrm{ads}}_{\textrm{a}}$[H2]}
    & \multicolumn{1}{r}{$\mathrm{E}^{\mathrm{ads}}_{\textrm{a}}$[H$2^{\ast}$]}
    & \multicolumn{1}{r}{$\mathrm{E}^{\mathrm{des}}_{\textrm{a}}$}
    & \multicolumn{1}{r}{$\mathrm{E}_{\mathrm{rxn}}$} \\ \hline
    LDA      & $0.60$ & $0.92$ & $2.23$ & $1.62$ \\
    PBE      & $0.34$ & $0.53$ & $1.95$ & $1.61$ \\
    PBE@HF   & $0.46$ & $0.50$ & $2.13$ & $1.76$ \\
    SCAN     & $0.35$ & $0.49$ & $2.32$ & $1.97$ \\
    B3LYP    & $0.56$ & $0.72$ & $2.50$ & $1.94$ \\
    PBE0     & $0.47$ & $0.51$ & $2.43$ & $1.96$ \\
    HSE06    & $0.46$ & $0.52$ & $2.38$ & $1.92$ \\
    RPA      & $0.66$ & $0.82$ & $2.16$ & $1.75$ \\
    HF       & $1.18$ & $1.02$ & $3.79$ & $2.62$ \\
    MP2      & $0.66$ & $0.78$ & $2.28$ & $1.62$ \\
    CCSD     & $0.79$ & $0.78$ & $2.92$ & $2.18$ \\
    CCSD(T)  & $0.70$ & $0.75$ & $2.62$ & $1.91$ \\
    QMC~\cite{filippi2002}  & $(0.09)0.63$ & $(0.05)0.75$ & $(0.09)2.91$ & $(0.05)2.20$ \\
    Expt.~\cite{duerr2001,hoefer1992,raschke2001} & $>0.6$ & $>0.6$ & $(0.10)2.50$ & $(0.30)1.90$ \\ \hline
    ZPE~\cite{steckel2001}  & $+0.09$      & $+0.09$      & $-0.11$      & $-0.20$      \\
  \end{tabular}
  \end{ruledtabular}
\end{table}

Finally, we examine the desorption mechanisms for the reaction. The
desorption barrier is defined as
$\mathrm{E}^{\mathrm{des}}_{\textrm{a}} =
\mathrm{E}_{\mathrm{transition}} - \mathrm{E}_{\mathrm{final}}.$ QMC
corrections using finite clusters~\cite{filippi2002} predict that none
of the H2 or H$2^{\ast}$ mechanisms are compatible with
temperature programmed desorption experiments~\cite{hoefer1992}, since
they yield too high desorption barriers for both mechanisms. Using
periodic CCSD(T), however, we find that desorption barriers for the
two mechanisms are very close and agree rather well with the
experimental estimate of $2.5\pm0.1$~eV. Furthermore, CCSD(T)
desorption energies are 2.62 and 2.67~eV for the H2 and H$2^{\ast}$
mechanisms. DFT-PBE predicts desorption energies of 1.95 and 2.14~eV,
respectively, vastly misjudging the absolute magnitude, as well as the
relative difference of the desorption barrier for the two
mechanisms. The SCAN functional improves the PBE desorption barriers,
however, only by ameliorating the description of the chemisorption
energy and not of the adsorption barrier. Hybrid functional results
are in satisfactory agreement with CSSD(T). Moreover, PBE0 and HSE06
estimated desorption energies are 2.43 and 2.38~eV for the H2
mechanism and 2.47 and 2.44~eV for the H$2^{\ast}$ one. We stress that
the two desorption energies are not only significantly higher than the
DFT-PBE ones but also not far from each other. B3LYP yields desorption
energies of 2.50 and 2.66~eV for the H2 and H$2^{\ast}$ mechanisms
respectively. Although the energies are close to the experimental
estimate we note that the overall picture for the reaction mechanism
is significantly different than the CCSD(T) one. The H2 pathway is
much prefered over the H$2^{\ast}$ one due to the mistreatment of the
relative difference of the two adsorption barriers, stemming from the
LDA part of the exchange-correlation functional. Finally the RPA
desorption energies, although they represent a significant improvement
over DFT-PBE, they still inherit shortcomings of the parent PBE
density functional, by favoring the H2 adsorption channel.

\section{Conclusion and Summary}

We have performed a range of DFT and quantum chemical
wavefunction based calculations for the H2 and H$2^{\ast}$
adsorption/desorption mechanisms of H$_2$ on the Si(100) surface at
low coverage. We show that periodic CCSD(T) calculations yield
excellent agreement with experimental results for the adsorption
barriers and the reaction energy.  In contrast to previous
calculations, we find similar activation energies for the H2 and
H$2^{\ast}$ adsorption mechanisms.  DFT-GGA and DFT-meta-GGA
functionals over-stabilize the H2 adsorption mechanism due to
incorrect ground state densities caused by self-interaction errors.
We argue that both a correct description of the H$_2$ molecule
dissociation, as well as of the surface dimer reconstruction is
essential for a precise interpretation of the reaction mechanisms.  We
note that hybrid functionals, like PBE0 and HSE06 slightly
underestimate the adsorption barriers, however, they yield adequate
results for the energetics of the reaction.  We have demonstrated that
high level periodic wavefunction based methods have the potential to
serve as accurate benchmark theories for predicting reaction
energetics on periodic surfaces, which will ultimately help to further
improve upon computationally more efficient yet less accurate methods.

\section{Acknowledgements}

This project has received funding from the European Research Council
(ERC) under the European Union’s Horizon 2020 research and innovation
program (grant agreement No 715594).  The computational results
presented here were conducted on the IBM iDataPlex HPC system HYDRA of
the Max Planck Computing and Data Facility (MPCDF).

\bibliographystyle{apsrev4-1}
\bibliography{bibliography}

\end{document}